\documentclass[conference]{IEEEtran} 
\usepackage{amsmath}
\usepackage{amsmath,amssymb,amsfonts,amsthm}
\usepackage{cite}
\usepackage{mathtools}
\usepackage{algorithm}                % Algorithm floating environment
\usepackage{bm}   
\usepackage[binary-units=true]{siunitx}
\DeclareSIUnit{\belmilliwatt}{Bm}
\newtheorem{theorem}{Theorem}

\usepackage{graphicx}
\usepackage{dsfont}
\usepackage{subcaption}
\usepackage{comment}

\usepackage{textcomp}
\usepackage[export]{adjustbox}
\usepackage{float}
\usepackage{booktabs}
\usepackage{multicol}
\usepackage{xparse}
\usepackage[left=1.62cm,right=1.62cm,top=1.9cm]{geometry}
\usepackage{color,soul}
\usepackage[bottom]{footmisc}
\usepackage[binary-units=true]{siunitx}
\DeclareSIUnit{\belmilliwatt}{Bm}
\DeclareSIUnit{\dBm}{\deci\belmilliwatt}
\DeclareSIUnit[per-mode=symbol,per-symbol=p]{\Bps}{\byte\per\second}

\sethlcolor{yellow}
\usepackage{algpseudocode}

\usepackage{hyperref}

\makeatletter
\def\BState{\State\hskip-\ALG@thistlm}
\makeatother

\IEEEoverridecommandlockouts%\IEEEpubid{\makebox[\columnwidth]{ 978-1-6654-6483-3/23/\$31.00 ©2023 IEEE \hfill} \hspace{\columnsep}\makebox[\columnwidth]{ }}
\begin{document}

    \title{ Toward Efficient Deployment and Synchronization in Digital Twins-Empowered Networks}

\author{
\IEEEauthorblockN{Hossam Farag and \v{C}edomir Stefanovi\'{c}}
Department of Electronic Systems, Aalborg University, Denmark\\
Email: hmf@es.aau.dk, cs@es.aau.dk
}

	\maketitle
	
%%%%%%%%%%%%%%%%%%%%%%%%%%%%%%%%%%%%%%%%%%%%
%%%%%%%%%%%%%%%%%%%%%%%%%%%%%%%%%%%%%%%%%%%%
	\begin{abstract}
Digital twins (DTs) are envisioned as a key enabler of the cyber–physical continuum in future wireless networks. However, efficient deployment and synchronization of DTs in dynamic multi-access edge computing (MEC) environments remains challenging due to time-varying communication and computational resources. This paper investigates the joint optimization of DT deployment and synchronization in dynamic MEC environments. A deep reinforcement learning (DRL) framework is proposed for adaptive DT placement and association to minimize interaction latency between physical and digital entities. To ensure semantic freshness, an update scheduling policy is further designed to minimize the long-term weighted sum of the Age of Changed Information (AoCI) and the update cost. A relative policy iteration algorithm with a threshold-based structure is developed to derive the optimal policy. Simulation results show that the proposed methods achieve lower latency, enhanced information freshness, and reduced system cost compared with benchmark schemes.

%  $\mathcal{O}_m$
	\end{abstract}
%%%%%%%%%%%%%%%%%%%%%%%%%%%%%%%%%%%%%%%%%%%%
	    \vspace{-1mm}
%%%%%%%%%%%%%%%%%%%%%%%%%%%%%%%%%%%%%%%%%%%%
\section{Introduction and Motivation}\label{sec:intro}
\vspace{-1mm}
%%%%%%%%%%%%%%%%%%%%%%%%%%%%%%%%%%%%%%%%%%%%
Digital Twins (DTs) are envisioned as a key enabler in 6G  for intelligent applications, providing the technological foundation for a seamless cyber–physical continuum that bridges the physical world with its digital counterpart~\cite{DT-surv2}.  A DT object can be designed to emulate a single entity (e.g., an edge server, cloud server, end device, or wireless channel), a single service (e.g., augmented reality), or even multiple services simultaneously, enabling optimal resource management and quality of service. The realization of DTs in wireless networks has attracted significant attention, particularly through their integration with multi-access edge computing (MEC) servers~\cite{DT-MEC1, DT-MEC2}. For instance, DT systems are expected to play a pivotal role in non-terrestrial networks\cite{DT-NTN2,DT-surv}, supporting emulation of satellite constellations, physical layer performance analysis, and link budget calculations.

The integration of DTs with MEC-enabled networks introduces the open challenge of efficiently placing and managing DT objects in dynamic edge networks. Variations in channel conditions and the available computational resources can hinder the efficiency of the DT from improving the performance of its physical counterpart, highlighting the need for adaptive mapping strategies between physical devices, i.e., physical twins (PTs), and their corresponding DTs. Current studies have explored a variety of algorithms with diverse optimization objectives, including data security~\cite{MEC1},  load balancing~\cite{MEC2}, latency reduction~\cite{MEC3}, energy efficiency~\cite{MEC5}, and multi-objective optimization~\cite{MEC4}. These works have investigated the use of DTs as a key enabling technology for improving the performance of MEC-empowered networks. However, designing efficient construction and maintenance mechanisms for DTs in dynamic MEC-empowered networks remains a largely unexplored problem. Moreover, each DT needs to continuously synchronize with its PT in real-time to keep its data fresh.
Age of Information (AoI)~\cite{AoI-def}, defined as the time elapsed since the most recent data update at a destination, has been widely used to measure DT consistency and determine state estimation accuracy\cite{AoI-DT1, AoI-DT2, AoI-DT3}. Although AoI captures temporal staleness, it disregards the content carried by the updates and the current knowledge at the receiver, i.e.,  fails to indicate  whether the delivered update contributes meaningful  information for maintaining the fidelity of the digital twin. In  many practical scenarios, consecutive updates may report unchanged physical states. In such cases, AoI is reset at every  successful reception, even if no new information has been  conveyed. This can result in misleadingly low AoI values that  overestimate the freshness of the DT. In contrast, the Age of Changed Information (AoCI)~\cite{AoCI} has been introduced to capture semantic freshness by distinguishing informative updates from redundant ones, which is particularly relevant for DT systems, where frequent transmissions may not always carry new state information. Unlike AoI, the AoCI decreases  only when a newly received update conveys changed information (new content) that modifies the DT state. If the update is redundant, the AoCI continues to grow, indicating that the DT has not improved its semantic consistency with the physical entity. In practice, a status change can only be detected when an update generated after the change is successfully received. Since the exact change instant is unknown without continuous monitoring, designing an update policy requires balancing freshness against energy consumption. Higher sampling rates improve timeliness but incur higher energy costs, while lower rates increase staleness and the risk of missed detections. The challenge is further aggravated by unreliable wireless channels, where packet losses may cause the receiver to incorrectly assume no state change has occurred. In this context, optimizing DT deployment and update scheduling is crucial to achieving energy-efficient and high-fidelity DT-enabled MEC networks. 

Motivated by the aforementioned challenges, in this paper, we first develop a deep reinforcement learning (DRL)-based algorithm for adaptive DT deployment. The proposed DRL framework decides the optimal DT placement strategy with the goal to minimize the interaction latency between the DT and its PT considering the dynamic network states. Second, we formulate a status update problem with the goal of minimizing the weighted sum of the AoCI and the DT update cost. We provide a lightweight relative policy iteration algorithm to find the optimal updating policy with low complexity. Performance evaluations have demonstrated the effectiveness of our proposed algorithms to achieve lower latency, improved information freshness, and reduced system cost compared to benchmark schemes.

%In that context, several variants of the AoI metric have been introduced, including sampling (SA)age~\cite{SAoI}, the age of incorrect information (AoII)~\cite{AoII} and age of synchronization (AoS)\cite{AoS}. While these variants enrich the notion of freshness, they cannot be directly computed at the destination. Specifically, SA requires knowledge of the precise sampling instant of the physical process, AoS depends on the generation time of the first update following the last synchronization, and AoII necessitates continuous access to the ground-truth state of the process. In contrast, 
The rest of the paper is organized as follows. Section~\ref{system-model} describes the system model. Section~\ref{deployment} introduces the DT optimization problem and Section~\ref{scheduling} presents the update scheduling problem. Performance evaluation are given in Section~\ref{results}, and finally
Section~\ref{sec:conclusions} concludes the paper.

%%%%%%%%%%%%%%%%%%%%%%5%%%%
\section{System Model}
\label{system-model}
\begin{figure}[t!] 
		\centering
		\includegraphics[width= 1\linewidth]{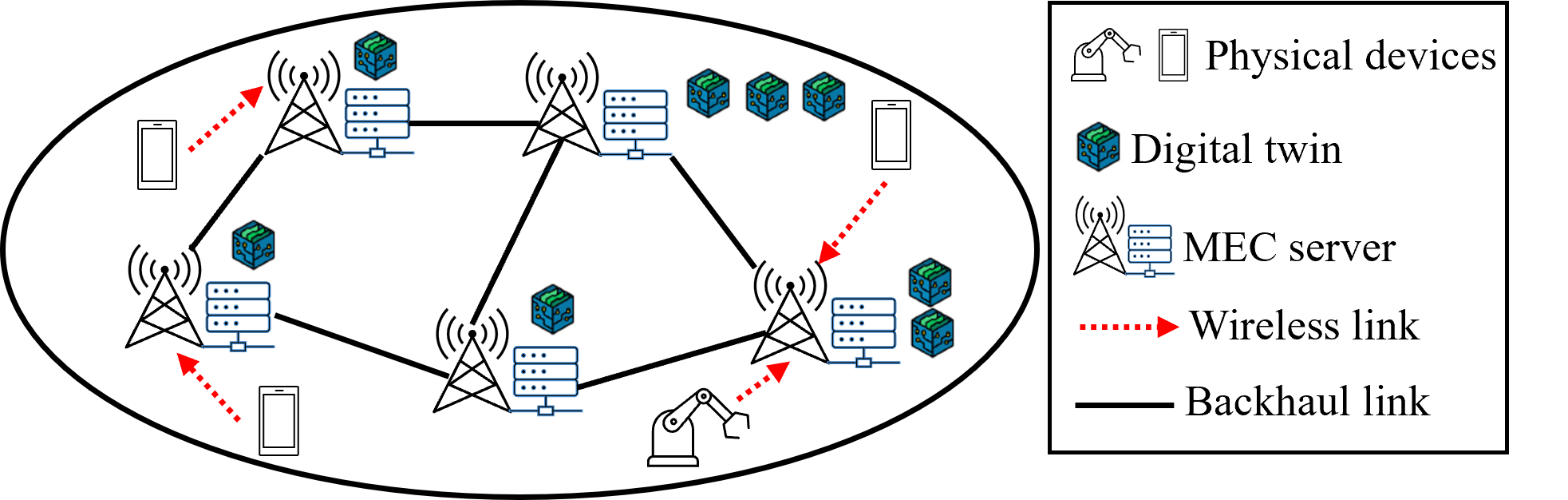} 
        %\vspace{-2mm}
		\caption{Illustration of DT-empowered MEC network.\label{network}}
		\vspace{-6mm}
	\end{figure}

\vspace{-2mm}
%%%%%%%%%%%%%%%%%%%%%%5%%%%
We consider DT-empowered network as depicted by Fig.~\ref{network} comprising a set $\mathcal{K}$ of $K$ devices (PTs), indexed by $k = \{1,2,\dots,K\}$. The system operates in discrete time slots, denoted by $\mathcal{T} = \{1,2,\dots,T\}$, each indexed by $t$ and has a fixed duration of $T_s$. The DT serves as a comprehensive digital counterpart of the physical device, capturing its hardware specifications, past operational records, and current status. 
The network consists of a set $\mathcal{B}$ of Base Stations (BSs), indexed by $b = \{1,2,\dots,B\}$, each integrated with a MEC server. All devices (PTs) transmit their updates to their associated BSs to synchronize with their corresponding DTs. For each BS, at most one device is scheduled to transmit to it at one time slot, and at most $B$ devices are scheduled to transmit simultaneously to $B$ BSs in one time slot. When a device $k$ is scheduled to update its DT, it generates a fresh data sample and sends it to the corresponding BS.  Let $a^t_{k,b}$ be a binary scheduling indicator, where  $a^t_{k,b} = 1$ if device $k$ is scheduled to transmit an update to BS $k$  in time slot $t$, and $a^t_{k,b} = 0$, otherwise. The wireless channel between device $k$ and BS $b$ is characterized by the channel power gain $h_{k,b}$. Let $d^t_{k,b}\in\{0,1\}$ denote the indicator of the outcome of the transmission of an update from device $k$ to its DT server in time slot $t$, where we have $d^t_{k,b}=1$ for a successful transmission and $d^t_{k,b}=0$, otherwise. We assume that failed updates will be discarded, and a new status update will be generated when $a^t_{k,b} = 1$.  
%%%%%%%%%%%%%%%%%%%%%%5%%%%%%%%%%%%%%%%%%%%%%%%%%5%%%%%%%%%%%%%%%%%%%%%%%%%%5%%%%
%%%%%%%%%%%%%%%%%%%%%%%%%%%%%%%%%%%%%%%%%%%%%%%5%%%%%%%%%%%%%%%%%%%%%%%%%%5%%%%%%%%%%%%%%%%%%%%%%%%%%5%%%%
%%%%%%%%%%%%%%%%%%%%%%%%%%%%%%%%%%%%%%%%%%%%%%%%%%%%%%%%
\subsection{DT interaction Latency}
%%%%%%%%%%%%%%%%%%%%%%%%%%%%%%%%%%%%%%%%%%%%%%%%%%%%%%%%
If device $k$ is scheduled to communicate with BS $b$ at slot $t$, its achievable transmission rate is given by
%%%%%%%%%%%%%%%%%%%%%%%%%%%%%%%%%%%%%%%%%%%%%%%%%%%%%%%%
\begin{equation} 
R_{k,s} = W \log \left( 1 + \frac{h_{k,b} p_{k,b}}{\sigma^2} \right),
\label{eq:rate}
\end{equation}
%%%%%%%%%%%%%%%%%%%%%%%%%%%%%%%%%%%%%%%%%%%%%%%%%%%%%%%%
where $W$ is the channel bandwidth, $p_{k,b}$ is the transmission power, and $\sigma^2$ denotes the noise power. Initially, the DT of device $k$ is constructed by transmitting its historical running data ($\mathcal{G}_k$) to the DT server via its associated BS. Denote the size of $\mathcal{G}_k$ as $D_k'$, then the transmission latency of $\mathcal{G}_k$ to the associated BS is given as    $T_{k,b}^{BS} = \frac{D_k'}{R_{k,b}}$.
The BS then forwards $\mathcal{G}_k$ to the DT server via backhaul (wired) link. Let $\beta$ denotes the latency required for transmitting one unit data in each unit distance and $d(b,m)$ is the distance between the BS $b$ and the DT server $m$. Then, the backhaul transmission latency is $   T_{b,m}^{back} = \beta D_k' d(b,m).$
The computation resources $F_m$ at each DT server are allocated to different devices for constructing their DTs. Let $f_{k,m}$ be  the assigned computational resources for processing the DT model, then the DT computation time $T_{k,m}^{comp}$ is given as $   T_{k,m}^{comp} = \frac{D_k'}{f_{k,m}}.$
Hence, the total DT construction latency $  T^{const}$ of the DT model at server $m$ can be expressed as $  T^{const}(k,m) = T_{k,b}^{BS}+T_{b,m}^{back}+T_{k,m}^{comp}.$ After constructing the DT, the PT $k$ keeps interacting with its DT where the PT transmits status updates of size $\Delta D_k$ to its associated DT server. The latency for updating the DT hosted by server $m$ is $T^{upd}(k,m)=\frac{\Delta D_k}{R_{k,s}}+\beta \Delta D_k d(b,m) + \frac{\Delta D_k}{f_{k,m}}.$
Therefore, the total interaction latency between a PT $k$ and its DT at server $m$ is $T^{inter}(k,m)=T^{const}(k,m)+T^{upd}(k,m)$.
%%%%%%%%%%%%%%%%%%%%%%%%%%%%%%%%%%%%%%%%%%%%%%%%%%%%%%%%

%%%%%%%%%%%%%%%%%%%%%%%%%%%%%%%%%%%%%%%%%%%%%%%%%%%%%%%%
\subsection{AoCI of DTs}
\vspace{-1mm}
%%%%%%%%%%%%%%%%%%%%%%%%%%%%%%%%%%%%%%%%%%%%%%%%%%%%%%%%

Next, we introduce the evaluation of the AoCI considering the DT of an arbitrary physical device. A physical device transmits a status update to its associated server to synchronize its DT counterpart aligning with the beginning of each time slot $t$. If the transmission is successful, the corresponding DT receives the update at the end of the slot. The AoI at a DT server represents the time elapsed since the generation time of the most recently received update. At time slot $t$, we denote $U(t) = \max \{ g_i \mid d_i \leq t \}$ as the generation time of the latest delivered update by the DT server, where $g_i$ and $d_i$ represent the generation and delivery time instants of update $i$, respectively. Denoting $\delta_t^k$ as  the instantaneous AoI of at the beginning of slot $t$, then we have $\delta_t^k = t - U(t)$.
AoI only captures temporal staleness, where the value of $\delta_t^k $ resets upon every successfully received update. As mentioned earlier, this could lead to redundant update transmission and waste of network resources when a newly received update does not modify the state of the DT. To overcome this limitation, we adopt the AoCI metric, which accounts for both the timing  and the content of updates.   Let $n(t) = \max \{ i \mid d_i \leq t \}$ denote the index of 
the most recent update received by the digital twin up to time slot $t$. Let $Y_j$ represent the physical state encapsulated in  update $j$, so that $Y_{n(t)}$ corresponds to the most recent state delivered to the twin. Define $q(t) = \max \{ j \mid Y_j \neq Y_{n(t)}, \, d_j \leq d_{n(t)} \}$ as the index of the last successfully received update that contained different information from $n(t)$. Note that all the update packets that have been
successfully received after $q(t)$ have the same content as the latest one received. Then, AoCI at the beginning of slot $t$ is given by
\vspace{-1mm}
%%%%%%%%%%%%%%%%%%%%%%%%%%%%%%%%%%
\begin{equation}
\Delta_t^k = t - U'(t),
\label{eq:aoci}
\vspace{-2mm}
\end{equation} 
%%%%%%%%%%%%%%%%%%%%%%%%%%%%%%%%%%
where $U'(t) = \min \{ g_s \mid d_{q(t)} < d_s \leq d_{n(t)} \}$ denotes the generation time of the first successful update received after $q(t)$. We next account for the randomness of content changes of the physical twin. Let $c^t_{k,b}\in \{0,1\}$ denote an indicator variable specifying whether the newly received update in slot $t$ contains information different from the previously received update. If $c^t_{k,b} = 1$, the content has changed; otherwise, $c^t_{k,b} = 0$. The probability of a content-changing update can then be written as
%%%%%%%%%%%%%%%%%%%%%%%%%%%%%%%%%%\
\begin{equation}
\Pr(c^t_{k,b} = 1) 
\stackrel{(a)}{=} \Pr \big( Y_{n(t)} \neq Y_{n(t)-1} \big) 
\stackrel{(b)}{=} 1 - p_{\delta_t^k},
\label{eq:content_prob}
\end{equation}
%%%%%%%%%%%%%%%%%%%%%%%%%%%%%%%%%%\begin{equation}
where $p_{\delta_t^k} = \Pr \big( X_{U(t)} = X_{U(t)-\delta_t^k} \big)$ is the return probability of the state of the PT, i.e., the likelihood that the PT remains in the same state over an interval of $\delta_t^k$ slots. Equality $(a)$ follows directly from 
the definition of $c^t_{k,b}$, while $(b)$ exploits the mapping between physical states $X_{U(t)}$ and the sampled updates $Y_{n(t)}$. If an update is successfully received 
($a^t_{k,b} = 1$, $d^t_{k,b} = 1$) and the physical state has changed 
($c^t_{k,b} = 1$), $\Delta_t^k$ resets to one; otherwise, $\Delta_t^k$
increases linearly. Hence, the evaluation of $\Delta_t^k$ can be expressed as
%%%%%%%%%%%%%%%%%%%%%%%%%%%%%%%%%
\begin{equation}
\Delta_{t+1}^k =
\begin{cases}
1, & a^t_{k,b} = 1,\, d^t_{k,b} = 1,\, c^t_{k,b} = 1, \\
\Delta_t^k + 1, & \text{otherwise}.
\end{cases}
\label{eq:aoci_dynamics}
\vspace{-2mm}
\end{equation}
%%%%%%%%%%%%%%%%%%%%%%%%%%%%%%%%%
We define $\hat{\delta}$ and $\hat{\Delta}$ as the upper bounds for AoI and AoCI, respectively. In line with the operational phases of digital twins, our problem can be decomposed into the following subproblems: i) DT placement, where the objective is to determine the optimal server locations and establish appropriate association strategies between servers and PTs,  and ii) scheduling, where servers schedules the transmission of PTs to keep synchronized.

%%%%%%%%%%%%%%%%%%%%%%%%%%%%%%%%%
\section{Optimal DT Deployment} \label{deployment}
%%%%%%%%%%%%%%%%%%%%%%%%%%%%%%%%%
%%%%%%%%%%%%%%%%%%%%%%%%%%%%%%%%%
\subsection{DT Placement Problem}
\vspace{-1mm}
%%%%%%%%%%%%%%%%%%%%%%%%%%%%%%%%%
Constructing DTs in each BS (MEC server) would incur a considerable transmission load, computation overhead, and energy consumption due to intensive resource consumption for maintaining DTs. Therefore, our network models considers a subset $\mathcal{M}\subset \mathcal{S}$ of the BSs as DT servers. We consider the association matrix $\boldsymbol{V}^{K\times B}=[v_{k,m}]$, where $v_{k,m}\in\{0,1\}$ is a binary indicator that is; $v_{k,m}=1$ if the DT of $k$ is maintained by server $m$, and  $v_{k,m}=0$, otherwise. Each device maintains exactly one DT instance, i.e., $\sum_{m\in \mathcal{S}} v_{k,m} = 1, \quad \forall k \in \mathcal{K}$. The DT servers may maintain the DTs of multiple devices with each DT server has a limited capacity of hosting a maximum of $N_m$ DTs, which is determined by the server resources. 

The objective of the DT placement strategy is to minimize the interaction latency between the PT and its DT. Denote $pl= \{\mathrm{loc}_m \mid m \in [1,B]\}$ as the placement policy of DT server, where $\mathrm{loc}_m$ represents the location of the DT server $m$ (deployed at which BS/server).  The average DT construction latency can be expressed as
%%%%%%%%%%%%%%%%%%%%%%%%%%%%%%%%%
\begin{equation}\label{const_latency}
  T^{inter}(pl, \boldsymbol{B}) = \frac{1}{KB} \sum_{k \in \mathcal{K}, m \in \mathcal{B}} T^{inter}(k,m) v_{k,m},
  \vspace{-2mm}
\end{equation}
%%%%%%%%%%%%%%%%%%%%%%%%%%%%%%%%%
where $T^{inter}(k,m)$ is the DT interaction latency considering the placement policy $pl$.  From \eqref{const_latency}, we can see that the DT construction latency is directly influenced by the placement policy ($pl$) and the association policy ($v_{k,m}\in\boldsymbol{V}$). Therefore, the DT placement and association problem can be formulated as 
\vspace{-3mm}
%%%%%%%%%%%%%%%%%%%%%%%%%%%%%%%%%%%%%%%%%%%%%%%%%%%%%%%%%%%%%%%%%%
\begin{align}
\textbf{P1:} \quad 
& \underset{pl,\boldsymbol{B}}{\text{min}} \quad 
 T^{inter}(pl, \boldsymbol{B})  \label{eq:opt_problem} \\
& \text{subject to:} \nonumber \\
& \quad \text{C1: } v_{k,m} \in \{0,1\}, 
\quad \forall k \in \mathcal{K}, \,m \in \mathcal{S}, \nonumber \\
& \quad \text{C2: } \sum_{k\in\mathcal{K}} v_{k,m} \leq N_m, 
\quad \forall m \in \mathcal{B},  \nonumber \\
& \quad \text{C3: } \sum_{m\in\mathcal{B}} v_{k,m} = 1, 
\quad \forall k \in \mathcal{K}, \nonumber \\
& \quad \text{C4: } \sum_{k\in\mathcal{K} }x_{x,m}f_{k,m}\leq F_m, 
\quad \forall k \in \mathcal{K}. \nonumber
\end{align}
%%%%%%%%%%%%%%%%%%%%%%%%%%%%%%%%%%%%%%%%%%%%%%%%%%%%%%%%%%%%%%%%%%
C1 enforces the binary nature of the association variable $v_{k,m}$, while C2 guarantees that maximum capacity of a DT server. C3 ensures that each device has a unique digital twin placement.  C4 indicates that the allocated computation resources does not exceed the total computational capacity of the server.
%%%%%%%%%%%%%%%%%%%%%%%%%%%%%%%%%
\subsection{DRL-based DT placement}
\vspace{-1mm}
%%%%%%%%%%%%%%%%%%%%%%%%%%%%%%%%%
We reformulate $\mathbf{P1}$ as a Markov Decision Process (MDP)  consisting of the quadruplet $\{\mathcal{S}(t), \mathcal{A}(t), R, \mathcal{S}(t+1)\}$, where $\mathcal{S}(t)$ represents the current system state, $\mathcal{A}(t)$ is the action taken, $R$ is the corresponding  reward, and $\mathcal{S}(t+1)$ denotes the next state. Below we describe the state space, action space, and reward design. \textbf{State Space}:  The system state includes the current placement of digital twin servers $p^l_j$, and the association decisions to the BSs $l_{ij}$, the active DTs at the server $J_b$, and the data rate between the users and the server $R_{k,b}$. $p^l_j\in\{0,1\}$ is a binary placement variable indicating whether BS $j$ hosts a DT server, and $l_{ij}\in\{0,1\}$ denotes whether user $i$ is connected to BS $j$. Hence, the system space is represented as $\mathcal{S}(t)\{p^l_j, l_{ij}, J_b, R_{k,b}\}$. \textbf{Action Space}: Each BS represents an agent that selects a joint action $a_j$ including (i) placement decision and (ii) association decision. DT placement decision is a binary decision $\{0,1\}$ where the agent decides whether to act as a DT server (1) or not (0). The association decision involves which devices associate with the BS for wireless access. Formally, the action space of such multi-agent system can be represented as $\mathcal{A}(t) = \left\{ a_{j} \; \middle| \; v_{ij} = (p^l_j, l_{ij}), \;\forall i \in \mathcal{K}, \;\forall j \in \mathcal{B} \right\}$. \textbf{Reward Function}: We design the reward function to ensure a balance between the interaction latency and the DT implementation cost. Let $\Theta$ denotes the average per-DT operational cost at the server, covering e.g., compute, memory, storage, power overhead. The reward function is given as $R=\beta R_T-(1-\beta)(R_E)$, where $R_T = -T^{inter}(pl, \boldsymbol{B})$ and $R_E=J_b\Theta$ and $\beta$ is the weight factor that decides the preference between latency and deployment cost. To derive the optimal placement and association strategies, we employ an  actor-critic deep reinforcement learning (DRL) framework. The aim is to maximize the expected discounted cumulative reward ,$\mathcal{R} = \sum_{t=1}^{T} \gamma^{t-1} R_t$, where $\gamma \in (0,1]$ is the discount factor. The actor network, parameterized by $\theta^\pi$, is responsible for generating actions,  while the critic network, parameterized by $\theta^Q$, evaluates the action quality. The action of agent $i$ at slot $t$ is $a_i(t) = \pi_i(s_t|\theta^\pi_i) + \mathcal{N},$ where $\mathcal{N}$ denotes exploration noise. The actor parameters are updated by
\begin{equation}
\theta^\pi \leftarrow \theta^\pi + \alpha \,
\mathbb{E}\!\left[
\nabla_{a_i} Q(s_t,a|\theta^Q) \,\middle|\,
a_i=\pi(s_t|\theta^\pi)
\right] \nabla_{\theta^\pi} \pi(s_t).
\end{equation}
%%%%%%%%%%%%%%%%%%%%%%%%%%%%%%%%%
where $y_t$ is the target value and $\alpha$ is the learning rate. Through this iterative actor--critic training, agents progressively refine their 
policies to minimize system latency while respecting computation and communication 
constraints. The use of shared system state information enables coordinated 
decisions among agents, ensuring convergence to efficient placement strategies 
for digital twin management.
%%%%%%%%%%%%%%%%%%%%%%%%%%%%%%%%%%%%%%%%%%%%%%%%%%%%%%%%%%%%%%%%%%
\begin{algorithm}[t]
\caption{Relative Policy Iteration}
\label{alg:relative}
\begin{algorithmic}[1]
\State \textbf{Initialization:} Set $j = 0$ and $\pi_{0}(s)=0$ for all states $s=(\Delta,\delta)\in\mathcal{S}$.
Select a reference state $s^{\dagger}$ and set $V_{0}(s^{\dagger})=0$.
\Repeat
    \Statex \hspace{1em}Given policy $\Psi_j(s)$, compute $\theta_j$ and $V_j(s)$ by solving the following $|\mathcal{S}|$ linear equations:
    \begin{equation*}
    \left\{
    \begin{aligned}
    \theta_j + V_j(s) &= C(s,\Psi_j(s))
        + \sum_{s'\in\mathcal{S}} P(s'|s,\Psi_j(s))\,V_j(s'), \\
    V_j(s^{\dagger}) &= 0.
    \end{aligned}
    \right.
    \end{equation*}
    \For{each $s=(\Delta,\delta)\in\mathcal{S}$}
        \If{
        $\displaystyle p_r(1)-p_r(\delta{+}1) 
        \le \frac{\delta}{V_j(\Delta{+}1,1)-V_j(1,1)}$
        \textbf{ and }
        $\displaystyle p_s(1{-}p_r(\delta))\Delta - p_s\delta - \omega C_k \ge 0$
        }
            \State $\pi_{j+1}(s)\leftarrow 1$
        \Else
            \State $\displaystyle 
            \Psi_{j+1}(s)\leftarrow 
            \arg\min_{a\in\mathcal{A}}
            \Big\{ C(s,a)
            + \sum_{s'\in\mathcal{S}} P(s'|s,a)V_j(s') \Big\}$
        \EndIf
    \EndFor
    \State $j\leftarrow j+1$
\Until{$\Psi_{j+1}(s)=\pi_j(s)$ for all $s\in\mathcal{S}$}
\State \textbf{Output:} Optimal policy $\Psi^\star \leftarrow \Psi_{j+1}$
\end{algorithmic}
\end{algorithm}
%%%%%%%%%%%%%%%%%%%%%%%%%%%%%%%%%%%%%%%%%%%%%%%%%%%%%%%%%%%%%%%%%%

%%%%%%%%%%%%%%%%%%%%%%%%%%%%%%%%%%%%%%%%%%%%%%%%%%%%%%%%%%%%%
%%%%%%%%%%%%%%%%%%%%%%%%%%%%%%%%%
\section{Update Scheduling Problem} \label{scheduling}
\vspace{-1mm}
%%%%%%%%%%%%%%%%%%%%%%%%%%%%%%%%%
The objective is to design an update scheduling policy that minimizes the total average cost, which is a weighted sum of the AoCI and the update cost. Hereafter, we focus on optimizing the scheduling policy for an arbitrary PT-DT pair, where the PT is already associated to specific BS $b$ (we omit the indices $b$ and $k$). Let  $\Psi = \{\boldsymbol{a}^t\}_{t \in \mathcal{T}}$ a stationary deterministic scheduling policy selected from the feasible set $\mathcal{P}$. The optimization problem can be expressed as
%%%%%%%%%%%%%%%%%%%%%%%%%%%%%%%%%
\begin{equation}
\min_{\Psi \in \mathcal{P}} 
\frac{1}{KT}\sum_{t=1}^{T} 
\left[\Delta_t + a_t\omega C \right],
\label{eq:cost}
\vspace{-1mm}
\end{equation}
%%%%%%%%%%%%%%%%%%%%%%%%%%%%%%%%%
where $C$ represents the update cost including the sampling cost and the transmission cost, where the cost for sampling is assumed to be negligible compared to that of transmission. $\omega$ is a non-negative weight balancing freshness and update cost.  The optimization problem in~\eqref{eq:cost} can be cast into an infinite-horizon 
average-cost Markov Decision Process (MDP) 
$(\mathcal{S},\mathcal{A}, \Pr(\cdot|\cdot,\cdot), C(\cdot,\cdot))$, 
where each element is described as follows. \textbf{States:} 
The state of the MDP at slot $t$ is defined by the tuple of the AoCI 
and the AoI, i.e., $s_t \triangleq (\Delta_t, \delta_t).$
Since both $\Delta_t$ and $\delta_t$ are bounded by their upper limits 
$\hat{\Delta}$ and $\hat{\delta}$, respectively, the state space 
$\mathcal{S}$ is finite. \textbf{Actions:} 
At each slot $t$, the action ($a_t$) is the scheduling decision, where $a_t=1$ indicates that the PT generates and transmits a status update to its DT, and $a_t=0$ means remaining idle. \textbf{Transition Probability:} Let $\Pr(s_{t+1}|s_t,a_t)$ denote the probability that the system 
moves from state $s_t$ to $s_{t+1}$ given the action $a_t$. 
Since the transmission success/failure event and the content-change 
event are independent, and according to the AoCI evolution dynamics 
in \eqref{eq:aoci_dynamics}. \textbf{Cost:} We define the instantaneous cost at slot $t$ as $C(s_t,a_t)=\Delta_t + a_t\omega C$. The above MDP is a finite-state, finite-action average-cost MDP. 
According to \cite[Theorem 8.4.5]{MDPP}, there exists a stationary 
deterministic average-optimal policy if the cost function is bounded 
and the induced Markov chain is unichain. In our case, the instantaneous 
cost is bounded, and the state $(\hat{\Delta},\hat{\delta})$ is reachable 
from all other states, ensuring a single recurrent class. 
Thus, there exists a stationary and deterministic optimal policy. The optimal policy $\Psi^\ast$ that minimizes the
average cost can be expressed as $\Psi^\ast(s)=\arg\min_{a\in\{0,1\}}\Big\{\, C(s,a)+\sum_{s'\in\mathcal{S}} \Pr(s'\!\mid s,a)\,V(s') \Big\}$,
where $V(s)$ is the value function. Closed-form expressions for $V(\cdot)$ are generally unavailable; classical dynamic programming methods (policy iteration, value iteration) are often used but face the curse of
dimensionality, hence are typically
computationally intensive. Before we introduce our proposed update policy, we define the following Theorem. 
%%%%%%%%%%%%%%%%%%%%%%%%%%%%%%%
%%%%%%%%%%%%%%%%%%%%%%%%%%%%%%%%%%%%%%%%%%%%%%%%%%%%%%%%%%%%%%
\begin{theorem}Given \(s=(\Delta,\delta)\), if $p_r(1)-p_r(\delta+1)\;\le\;\frac{\delta}{\,V_j(\Delta+1,1)-V_j(1,1)\,}$ for any \(\Delta\) and $j$~\(\in\mathbb{Z}_{\ge 0}\), then the optimal updating policy \(\Psi^\ast(s)\) exhibits a \(\Delta\)-threshold structure for each fixed \(\delta\). Specifically, the optimal action is to update whenever \(\Delta \ge \Delta(\delta)\) where $\Delta(\delta)$ is the minimum integer value that satisfies $p_s\big(1-p_r(\delta)\big)\Delta \;-\; p_s\,\delta \;-\; \omega C_k \;\ge\; 0$. 
\end{theorem}
The proof of Theorem 1 is omitted due to the limited space of the paper.
To efficiently obtain the optimal scheduling policy, we adopt 
relative policy iteration algorithm detailed by Algorithm~\ref{alg:relative}, which is based on Theorem 1. Directly applying conventional policy iteration can be computationally intensive due to the 
large state space involved. The proposed method reduces the 
computational burden by exploiting the characteristics of the 
state transitions and the cost function. Since the maximum values of the AoCI 
and AoI are $\hat{\Delta}$ and $\hat{\delta}$, respectively, the state 
space has cardinality $|\hat{\Delta}|\times|\hat{\delta}|$. The proposed 
relative policy iteration reduces the per-iteration complexity 
from $\mathcal{O}\!\left((|\hat{\Delta}|\times|\hat{\delta}|)^3\right)$ 
of standard methods to 
$\mathcal{O}\!\left((|\hat{\Delta}|\times|\hat{\delta}|)^2\right)$, 
which makes the algorithm scalable for practical scenarios. 
%%%%%%%%%%%%%%%%%%%%%%%%%%%%%%%
\section{Performance Evaluation}
\label{results}
  \vspace{-1mm}
%%%%%%%%%%%%%%%%%%%%%%%%%%%%%%%
We consider a network area of $1000\textrm{m} \times 1000\textrm{m}$ where the devices (PTs) and the MEC servers are randomly distributed. Over the simulation time, the channel state, the locations of the PTs and the computing resources of the MEC servers are varied. The channel between the PTs and the MEC server undergoes Rayleigh fading. The time slot duration is set to $10$~ms with a total of 1000 time slots considered. The final results are averaged over 1000 simulations runs. Table~\ref{t1} includes the simulation parameters adopted in this work. 
%%%%%%%%%%%%%%%%%%
\begin{table}[t!]
		\centering
		\caption{Simulation parameters}
                \vspace{-2mm}
		\label{t1}
		\begin{tabular}{ll}
			\toprule
			Parameter & Value \\
				\midrule
			Channel bandwidth (W) & 20 MHz\\
            Computation resources of MEC servers ($F_m$) & [5-20] GHz\\
            Noise power & $-174$ dBm/Hz\\
            Training iterations & $5000$ iterations\\
            Discount factor $\gamma$ & $0.6$ \\
			\bottomrule
		\end{tabular}	
        \vspace{-6mm}
	\end{table}
%%%%%%%%%%%%%%%%%%%%%%%%%%%%%%%%%%%%%%%%%%%%%%%%%%%%%%%%%%%%

\begin{figure}[t!] 
		\centering
		\includegraphics[width= 0.73\linewidth]{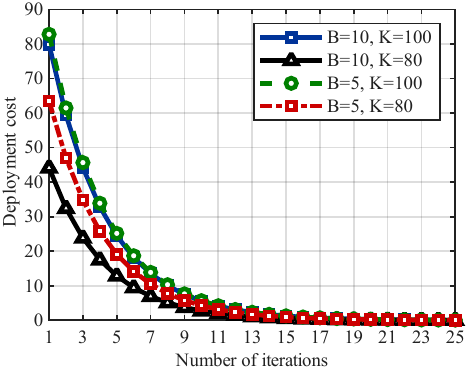} 
        \vspace{-2mm}
		\caption{Evaluation of the DT deployment cost with varying $N$ and $B$.\label{converg}}
		\vspace{-5mm}
	\end{figure}

Fig.~\ref{converg} shows the performance of the DRL-based optimization process for the DT deployment problem in~\eqref{eq:opt_problem}. The deployment cost represents a normalized cost metric that reflects  a combination of construction latency and DT operational cost. It can be observed that the deployment cost decreases rapidly and stabilizes after few iterations, indicating fast convergence of the proposed DRL-based algorithm. Also, the figure highlights that the number of PTs ($K$) has a more pronounced impact on the system cost compared to the number of MEC servers ($B$), since higher density of PTs intensifies competition for limited communication and computation resources. Conversely, increasing $B$ provides additional resources, thereby reducing the overall cost. These results demonstrate that the proposed scheme is able to efficiently adapt to varying network scales while maintaining low deployment cost and rapid convergence.
%%%%%%%%%%%%%%%%%%%%%%5%%%%%%%%%%%%%%%%%%%%%%%%%%5%%%%%%%%%%%%%%%%%%%%%%%%%%5%%%%
%%%%%%%%%%%%%%%%%%%%%%%%%%%%%%%%%%%%%%%%%%%%%%%5%%%%%%%%%%%%%%%%%%%%%%%%%%5%%%%%%%%%%%%%%%%%%%%%%%%%%5%%%%
%%%%%%%%%%%%%%%%%%%%%%5%%%%%%%%%%%%%%%%%%%%%%%%%%5%%%%%%%%%%%%%%%%%%%%%%%%%%5%%%%
    \begin{figure}[t!] 
		\centering
		\includegraphics[width= 0.8\linewidth]{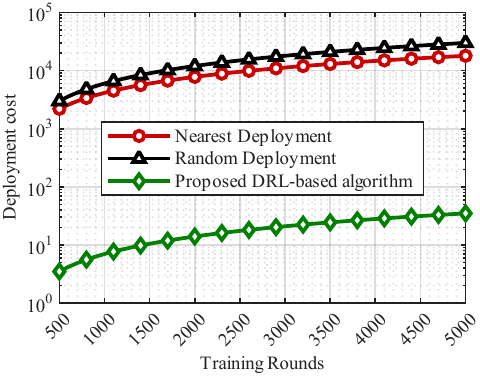} 
        \vspace{-3mm}
		\caption{Performance comparison against baselines methods.\label{compare1}}
		\vspace{-6mm}
	\end{figure}
%%%%%%%%%%%%%%%%%%%%%%%%%%%%%%%%%%%%%%%%%%%%%%%5%%%%%%%%%%%%%%%%%%%%%%%%%%5%%%%%%%%%%%%%%%%%%%%%%%%%%5%%%%
We further compare  our proposed deployment algorithm with two baseline methods as depicted by Fig.~\ref{compare1}. The first baseline is random deployment where the DTs are placed and associated randomly while the second one is the nearest deployment where the PTs associates with the closest MEC server to implement their DTs. As shown, the proposed method consistently achieves a markedly lower total system cost by dynamically adapting association decisions to instantaneous network states through the trained actor–critic model. This demonstrates that the proposed scheme not only reduces the overall time cost but also efficiently identifies near-optimal association strategies under varying network conditions.

\begin{figure}[t!] 
		\centering
		\includegraphics[width= 0.7\linewidth]{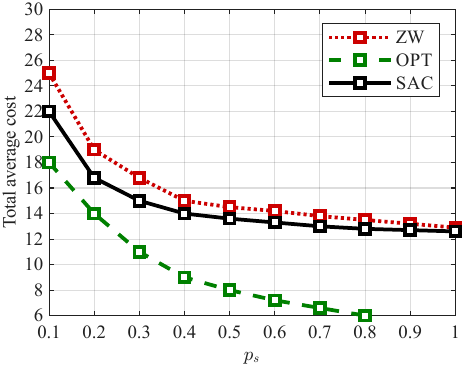} 
        \vspace{-2mm}
		\caption{Performance comparison against baselines methods with $C=12$ and $\omega=1$.\label{totalcost}}
		\vspace{-4mm}
	\end{figure}

\begin{figure}[t!] 
		\centering
		\includegraphics[width= 0.7\linewidth]{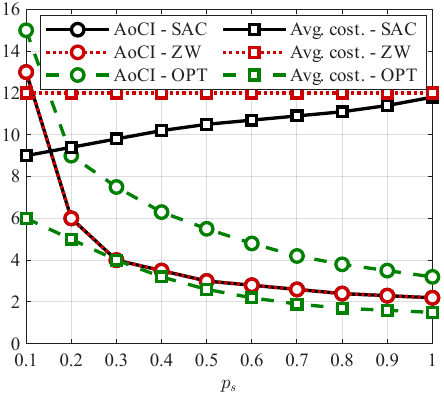} 
        \vspace{-2mm}
		\caption{Performance comparison against baselines deployment methods with $C=12$ and $\omega=1$.\label{compare}}
		\vspace{-7mm}
	\end{figure}

Next, we evaluate our proposed update scheduling policy and compare it against zero-wait policy (ZW) and sample-at-change policy (SAC). In the ZW policy, the sensor generates and transmits an update in every time slot, irrespective of state changes. In contrast, the SAC policy triggers a new update only when the current state differs from the last received one at the destination. The SAC policy serves as an ideal benchmark that achieves the minimum possible AoCI. Fig.~\ref{totalcost} confirms that our proposed scheduling policy outperforms both ZW and SAC policies with more cost reduction for higher values of $p_s$. The reason can be further illustrated by Fig.~\ref{compare} which illustrates the separate contributions of the average AoCI and the average update cost. The ZW and SAC policies achieve lower AoCI than the proposed policy, however, with a higher average update cost $\omega C_k$. Particularly, in our proposed update policy, the PT remains idle until the AoCI hits a certain threshold, hence increasing the average AoCI compared to ZW and SAC policies. In ZW, the PT generates and transmits updates at every time slot, incurring a constant and excessive update cost regardless of the transmission reliability and the status change. The update cost of the SAC policy increases with $p_s$ as it only focuses on the AoCI improvements and eventually degenerates to ZW policy when $p_s=1$. In contrast, our proposed update policy ensures a tradeoff between the AoCI and the update cost where the PT becomes reluctant to transmit updates when the AoCI is small to avoid excessive and unnecessary update cost.

%%%%%%%%%%%%%%%%%%%%%%5%%%%%%%%%%%%%%%%%%%%%%%%%%5%%%%%%%%%%%%%%%%%%%%%%%%%%5%%%%
%%%%%%%%%%%%%%%%%%%%%%%%%%%%%%%%%%%%%%%%%%%%%%%5%%%%%%%%%%%%%%%%%%%%%%%%%%5%%%%%%%%%%%%%%%%%%%%%%%%%%5%%%%

%%%%%%%%%%%%%%%%%%%%%%5%%%%%%%%%%%%%%%%%%%%%%%%%%5%%%%%%%%%%%%%%%%%%%%%%%%%%5%%%%
%%%%%%%%%%%%%%%%%%%%%%%%%%%%%%%%%%%%%%%%%%%%%%%5%%%%%%%%%%%%%%%%%%%%%%%%%%5%%%%%%%%%%%%%%%%%%%%%%%%%%5%%%%

%%%%%%%%%%%%%%%%%%%%%%%%%%%%%%%
\section{Conclusion}
\label{sec:conclusions}
\vspace{-1mm}
%%%%%%%%%%%%%%%%%%%%%%%%%%%
This paper proposed an integrated framework for DT deployment and synchronization in MEC-enabled networks. A DRL-based placement strategy was developed to minimize interaction latency, while a lightweight AoCI-aware update policy was introduced to balance information freshness and update cost. Numerical results confirmed the superior performance of the proposed methods in terms of latency reduction and energy efficiency. Future work will extend this approach to multi-agent and federated settings for large-scale DT orchestration in 6G networks.
%%%%%%%%%%%%%%%%%%%%%%%%%%%%
%\section*{Acknowledgement}
%%%%%%%%%%%%%%%%%%%%%%%%%%%
%This paper has received funding from the European Union’s Horizon 2020 research and innovation programme under Grant Agreement number 856967. 
	\bibliographystyle{IEEEtran}

\bibliography{mybib}

\end{document}